\documentclass[12pt,preprint]{aastex}

\slugcomment{}

\shorttitle{Filaments in Vela Jr. Remnant}
\shortauthors{Bamba et al.}

\begin{document}

\title{{\it Chandra} Observations of A Galactic Supernova Remnant Vela Jr.:
A New Sample of Thin Filaments Emitting Synchrotron X-Rays}

\author{
Aya Bamba\altaffilmark{1},
Ryo Yamazaki\altaffilmark{2},
Junko S.\ Hiraga\altaffilmark{3}
}
\altaffiltext{1}
{RIKEN (The Institute of Physical and Chemical Research)
2-1, Hirosawa, Wako, Saitama 351-0198, Japan
}
\altaffiltext{2}
{Department of Earth and Space Science,
Graduate School of Science, Osaka University,
Toyonaka, Osaka 560-0043, Japan
}
\altaffiltext{3}
{Department of High Energy Astrophysics
Institute of Space and Astronautical Science (ISAS)
Japan Aerospace Exploration Agency (JAXA)
3-1-1 Yoshinodai, Sagamihara, Kanagawa 229-8510, Japan
}

\email{bamba@crab.riken.jp,
ryo@vega.ess.sci.osaka-u.ac.jp,
jhiraga@astro.isas.jaxa.jp
}

\begin{abstract}
A galactic supernova remnant (SNR) Vela Jr. 
(RX~J0852.0$-$4622, G266.6$-$1.2)
shows sharp filamentary structure on the north-western edge of the remnant
in the hard X-ray band.
The filaments are so smooth
and located on the most outer side of the remnant.
We measured the averaged scale width of the filaments ($w_u$ and $w_d$)
with excellent spatial resolution of {\it Chandra},
which are
in the order of the size of the point spread function of {\it Chandra}
on the upstream side
and 49.5 (36.0--88.8)~arcsec on the downstream side,
respectively.
The spectra of the filaments are very hard and have no line-like structure,
and were well reproduced with an absorbed power-law model
with $\Gamma = $2.67 (2.55--2.77),
or a {\tt SRCUT} model with
$\nu_{rolloff}$ = 4.3 (3.4--5.3)$\times 10^{16}$~Hz
under the assumption of $p=0.3$.
These results imply that the hard X-rays are
synchrotron radiation emitted by accelerated electrons,
as mentioned previously.
Using a correlation between a function ${\cal B} \equiv \nu_{rolloff}/w_d^2$
and the SNR age,
we estimated the distance and the age of Vela Jr.:
the estimated distance and age are
0.33 (0.26--0.50)~kpc and 660 (420--1400)~years,
respectively.
These results are consistent with previous reports,
implying that ${\cal B}$--age relation may be a useful tool to
estimate the distance and the age of
synchrotron X-ray emitting SNRs.
\end{abstract}


\section{Introduction}
\label{sec:intro}

Supernova remnants (SNRs) play crucial roles
for heating and chemical evolution of galaxies.
Their shocks are also famous as cosmic-ray accelerators.
\citet{koyama1995} discovered synchrotron X-rays
from shells of SN~1006,
which is the first observational result
indicating that
SNRs accelerate electrons up to $\sim$TeV.
At present, several SNRs have been identified as 
synchrotron X-ray emitters
(e.g.\ RX~J1713.7$-$3946, \citealt{koyama1997,slane1999};
RCW~86, \citealt*{bamba2000,borkowski2001,rho2002};
Cas~A., \citealt{vink2003};
see also \citealt{bamba2005,bambaphd}).

The most plausible acceleration mechanism is
the diffusive shock acceleration (DSA)
\citep[e.g.][]{bell1978,drury1983,%
blandford1987,jones1991,malkov2001},
which can accelerate charged particles on the shock
into a power-law distribution,
similar to the observed spectrum of cosmic rays on the earth.
However, since there is still little observational information
and theoretical understanding,
there are many unresolved problems,
such as the injection efficiency,
the maximum energy of accelerated particles, 
the configurations of magnetic field, and so on.
The magnetic field amplification is also pointed out  
as an remarkable process \citep{bell2001,lucek2000}.
Recently,
it is found that
nonthermal X-rays from the shells of historical SNRs
concentrate on very narrow filamentary regions
\citep{bamba2003,bamba2005}.
Together with the roll-off frequency ($\nu_{rolloff}$) of synchrotron X-rays
\citep{reynolds1998,reynolds1999},
\citet{bamba2005} found that
an empirical function,
${\cal B} \equiv \nu_{rolloff}/w_d{}^{2}$,
decreases with the age of an SNR
($\cal B$---age relation).
The relation may reflect the time evolution of the magnetic field
around the shock front \citep{bamba2005}.
However, the $\cal B$---age relation has still many uncertainties
because of poor statistics.
Clearly other new samples with synchrotron X-ray filaments are desired.

Vela Jr. (RX~J0852.0$-$4622, G266.6$-$1.2) is 
one of the most curious galactic SNR
with large radius of about 60~arcmin
\citep{green2004}.
It was discovered by {\it ROSAT} \citep{achenbach1998}.
and well known 
that the {\it COMPTEL} team reported
a possible detection of $1.157$~MeV
$\gamma$-ray line emitted by $^{44}$Ti
\citep{iyudin1998}.
This fact strongly indicates that
Vela Jr. is a nearby and very young SNR 
because the half lifetime of $^{44}$Ti is only about 60~years,
however, there remains some doubt about the detection
\citep{shonfelder2000}.
\citet{tsunemi2000} reported the presence of Ca K line,
suggesting that there are a plenty of Ca produced from $^{44}$Ti.
On the other hand,
\citet{slane2001} reported that
the 1$\sigma$ upper-limit of Sc-K line is
$4.4\times 10^{-6}$~cm$^{-2}$s$^{-1}$.
Recently, \citet{iyudin2005} found the possible detection of
Ti and Sc K emission lines with {\it XMM-Newton}.
For the 78.4~keV line emission from $^{44}$Ti,
\citet{vonkeinlin2004} obtained
the upper-limit of $1.1\times 10^{-4}$~cm$^{-2}$s$^{-1}$
with {\it INTEGRAL}/SPI.
The presence of a central source has been also
suggested \citep{achenbach1998},
which was confirmed with {\it Chandra} \citep{pavlov2001}. 
The spectrum of the source, CXOU~J085201.4$-$461753,
together with the absence of any optical counter part \citep{mereghetti2002},
reminds us of a neutron star.
These observational reports might indicate that
the progenitor  is a core-collapse supernova (SN).

The other fact which makes this SNR famous is
the existence of nonthermal X-rays from the rims \citep{slane2001},
which implies that
the rims of Vela Jr. are cosmic ray accelerators
like SN~1006 \citep{koyama1995} and other SNRs with synchrotron X-rays
\citep[e.g.][and references therein]{bambaphd}.
Recently, TeV $\gamma$-rays with energies greater than 500~GeV
are also detected at the 6$\sigma$ level 
from the north-western rim of the SNR with {\it CANGAROO-II}
\citep{katagiri2005} and H.E.S.S.\ \citep{aharonian2005}.
These $\gamma$-rays are produced by accelerated electrons and/or
protons with energies more than $\sim$~TeV.

Despite such an interesting source,
the precise age of the remnant is still unknown.
There are some
spikes in nitrate concentration measured in an Antarctic ice core,
which might be regarded as signals of SNe \citep{burgess2000}.
One of them is not yet identified with historical SNe
ever known.
If the spike really associated with the progenitor of Vela Jr.,
the SN may have occurred around AD~1320 and the age of the remnant
is about 680~yrs.
However, there are many uncertainties, hence further studies are
necessary.
In addition, an unclear issue is
the distance to the remnant,
which brings us numerous information,
such as the total luminosity of the $^{44}$Ti emission line,
the average expansion velocity,
and so on.
It may also give us some answer on the current debate whether or not
Vela \citep[the distance of 250~pc;][]{cha1999} and Vela Jr.
interact with each other.
\citet{iyudin1998} estimated the distance to be $\sim$200~pc
with an assumption that the age is 680~years mentioned above,
while \citet{slane2001} suggested that the distance is 1--2~kpc
from the relatively large absorption column.
Although \citet{moriguchi2001} observed the molecular distribution 
using $^{12}$CO (J = 1---0) emission with NANTEN,
they could not determine the precise distance
because the remnant is located near 
the tangential point of a galactic arm.
The upper limit of the distance they estimated is $\sim$1~kpc.

In this paper, using the {\it Chandra} data 
for the first time,
we report on the analysis of hard X-ray filaments associated with
Vela~Jr. 
and discuss on the acceleration mechanism at the shock of the SNR,
its age, and the distance to the SNR.
This paper is organized as follows.
We summarize the observation details of Vela~Jr. with {\it Chandra}
in \S~\ref{sec:obs}.
Section~\ref{sec:results} describes
the results of the observations.
We  discuss on
$^{44}$Sc emission line
reported by \citet{iyudin2005} in
\S~\ref{sec:Sc-line},
 the origin of nonthermal X-rays in  \S~\ref{sec:discuss},
on the interaction between the SNR and molecular clouds (\S~\ref{sec:MC}),
and on the origin of TeV $\gamma$-rays (\S~\ref{sec:TeV}).
In \S~\ref{sec:estimation},
we argue the distance and the age of the SNR
using the correlation between a function $\cal B$ and the SNR age.

\section{Observations}
\label{sec:obs}

\begin{figure}[hbtp]
\epsscale{0.5}
\plotone{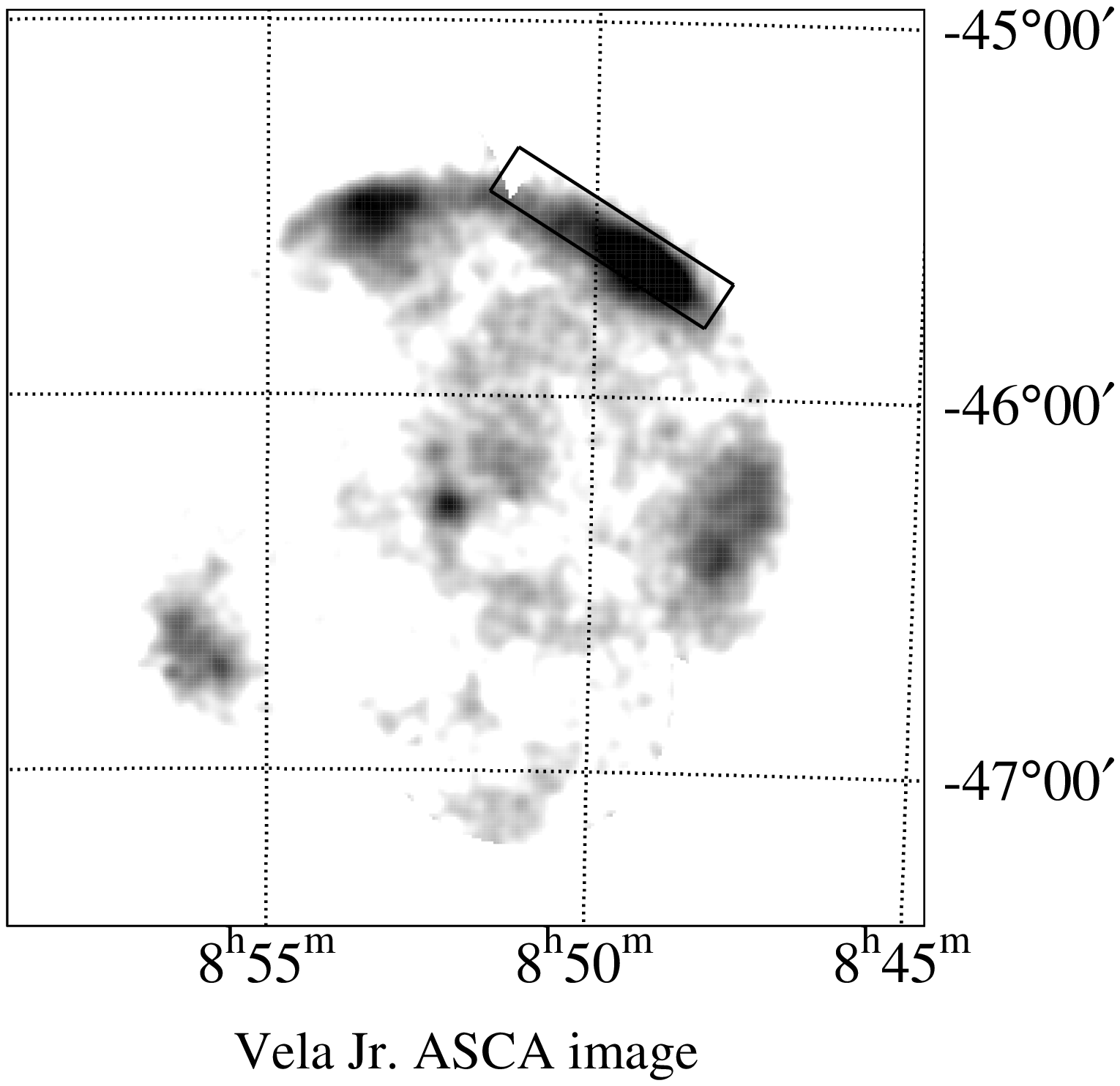}
\caption{{\it ASCA} GIS image of Vela Jr.
in the 0.7--10.0~keV band \citep{tsunemi2000}.
Gray-scale is in logarithmic and coordinates are in J2000.
The solid rectangle shows the field of view of {\it Chandra}
ACIS-S array.
}
\label{fig:ASCA}
\end{figure}

We used the {\it Chandra} archival data of the ACIS
of the north-western (NW) rim of Vela Jr. (ObsID = 3846 and 4414).
Figure~\ref{fig:ASCA} shows the {\it ASCA} GIS image \citep{tsunemi2000}
with the field of view of {\it Chandra} observations.
The satellite and the instrument are described
by \citet{weisskopf2002} and \citet{garmire2000}, respectively.
Data acquisition from the ACIS was made in the Timed-Exposure Faint mode.
The data reductions and analysis were made using the {\it Chandra}
Interactive Analysis of Observations (CIAO) software version 3.0.2.
Using the Level 2 processed events provided by the pipeline
processing at the {\it Chandra} X-ray Center,
we selected {\it ASCA} grades 0, 2, 3, 4, and 6, as the X-ray events.
The ``streak'' events on the CCD chip S4 were removed
by using the program {\it destreak}%
\footnote{See http://asc.harvard.edu/ciao2.3/ahelp/destreak.html.}
in CIAO.
In order to make statistics better,
we improved the astrometory of the data
following the CIAO data analysis threads
and combined these data.
The exposure time of each observation is
39~ks (ObsID = 3846) and 35~ks (ObsID = 4414), respectively.
Table~\ref{obs_log} summarizes about these observations.
Hereinafter,
all results are from the combined data
so far as there is no mention.

\section{Results}
\label{sec:results}

\begin{figure}[hbtp]
\epsscale{0.8}
\plotone{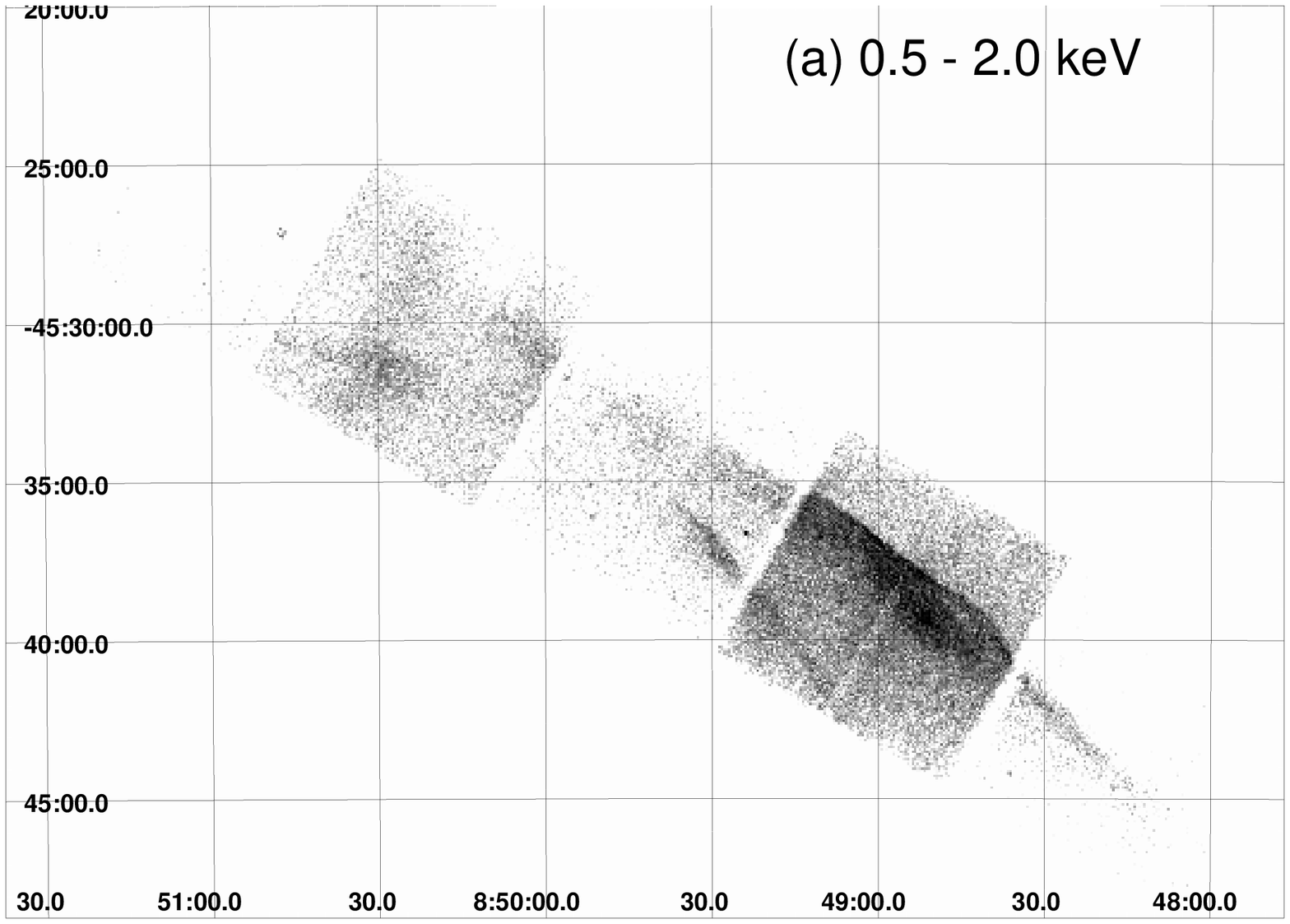}
\plotone{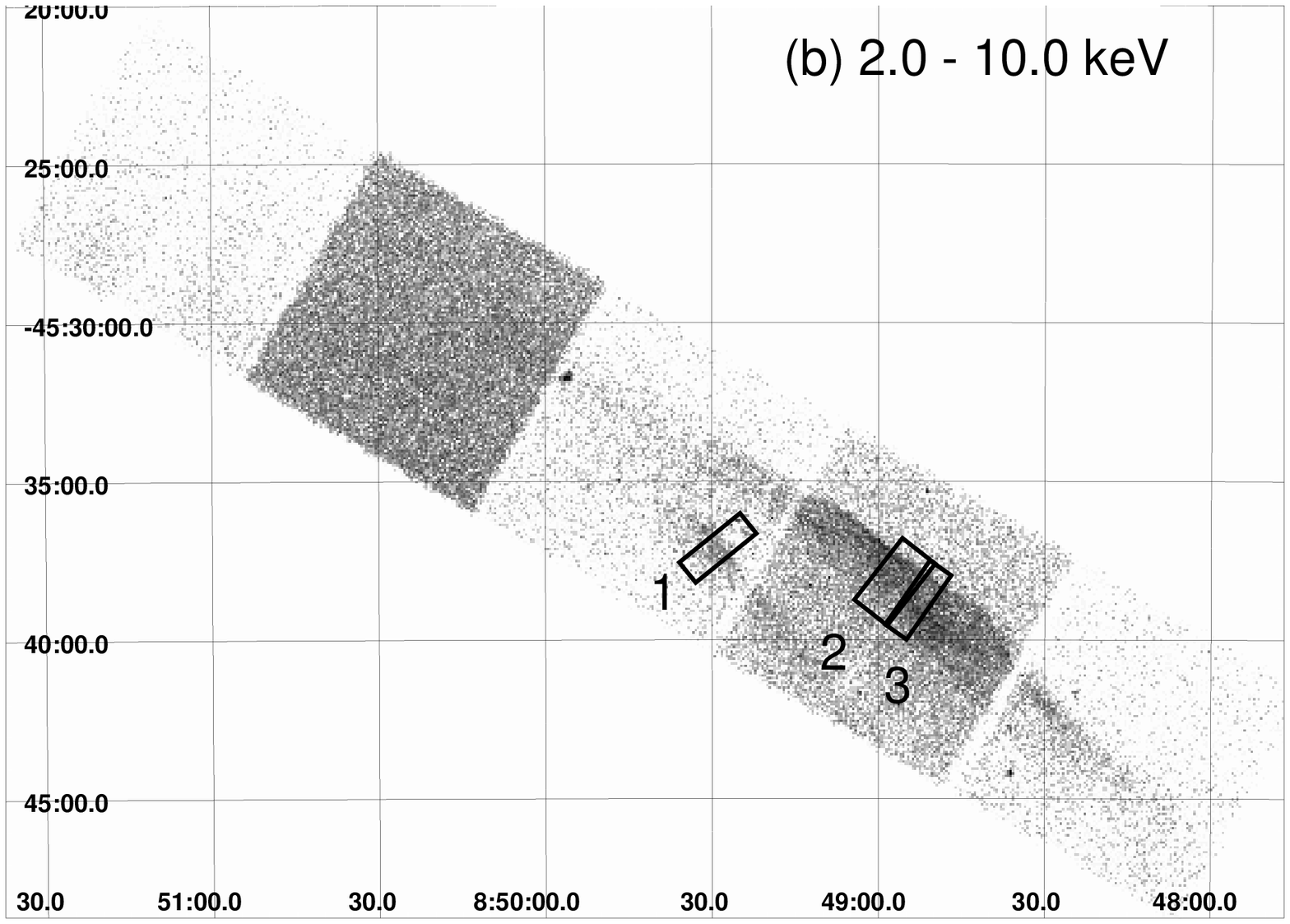}
\caption{{\it Chandra} images of the NW rim of Vela Jr.
in the 0.5--2.0~keV band (a) and 2.0--10.0~keV band (b).
Gray-scales are in logarithmic and coordinates are in J2000,
binned with 4~arcsec scale.
Regions to make profiles are shown in (b)
with solid rectangles.
}
\label{fig:images}
\end{figure}

Figure~\ref{fig:images} shows images of the NW rim
in the 0.5--2.0~keV (a) and 2.0--10.0~keV (b) bands,
binned with 4~arcsec scale.
Neither the subtraction of background photons, smoothing process,
nor correction of the exposure time did not be performed.
The difference of background level among the chips are caused by
that of the CCD chips (back-illuminated and front-illuminated).
These images look alike,
and show very straight filamentary structures clearly
on the outer edge of the rim.
Two filaments can be well recognized especially in the upper panel.
\citet{iyudin2005} suggests that 
the inner filament may be the reverse shock emission.
However, we have no clue to distinguish
whether it indicates the reverse shock or apparently overlaid forward 
shock via the projection effect.
The filaments are very similar to those in SN~1006
\citep{long2003,bamba2003},
indicating that
they may be efficient accelerators of electrons
like SN~1006 filaments.
Then 
we made analysis 
in the same way to our previous one
for SN~1006 \citep{bamba2003}.

For the first step,
we made a spectrum from the whole emitting region of this rim.
Since the rim is covered by two CCD chips (chip~6 and 7),
source spectra were made separately from each chip.
Background photons were accumulated from outer regions of the rim.
All the spectra
extend up to $>$2~keV,
indicating that 
the hard X-ray emission is nonthermal.
There is some line-like feature below 2~keV,
which probably comes from thermal emission of Vela and/or Vela Jr.
as already mentioned \citep{slane2001}.
Since our data does not have
sufficient statistics to examine the properties of thermal plasma,
we ignored the component.
In \citet{hiraga2005},
detailed analysis for the thermal plasma is done
with the {\it XMM-Newton} deep observation.
We fitted the spectra with an absorbed power-law model
only with photons above 2~keV,
in order to avoid the contamination of thermal photons.
The absorption column was subsequently calculated using the cross sections by
\citet{morrison1983} with the solar abundances \citep{anders1989}.
The best-fit parameters are shown in Table~\ref{total_spec}.

\begin{figure}[hbtp]
\epsscale{0.32}
\plotone{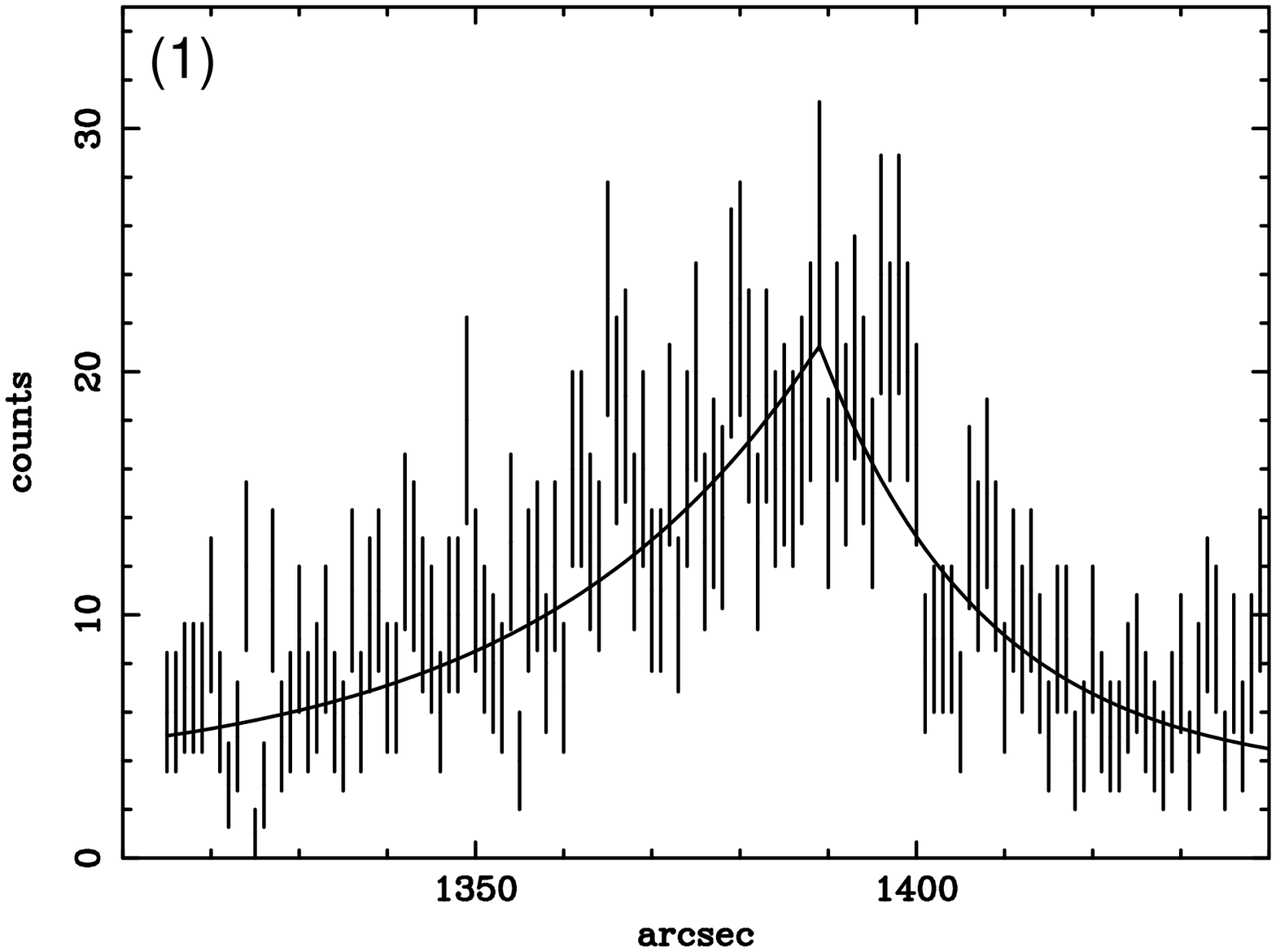}
\plotone{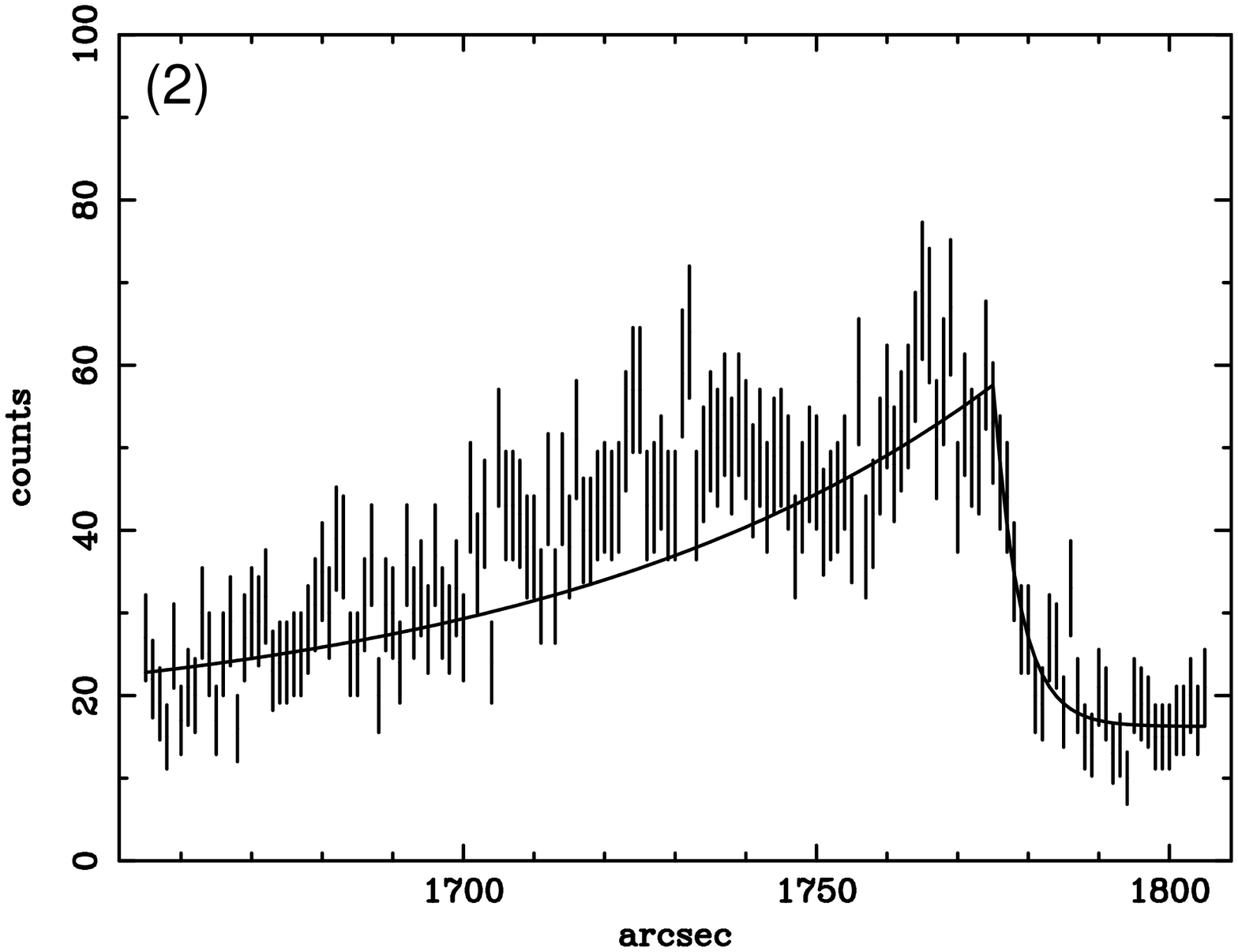}
\plotone{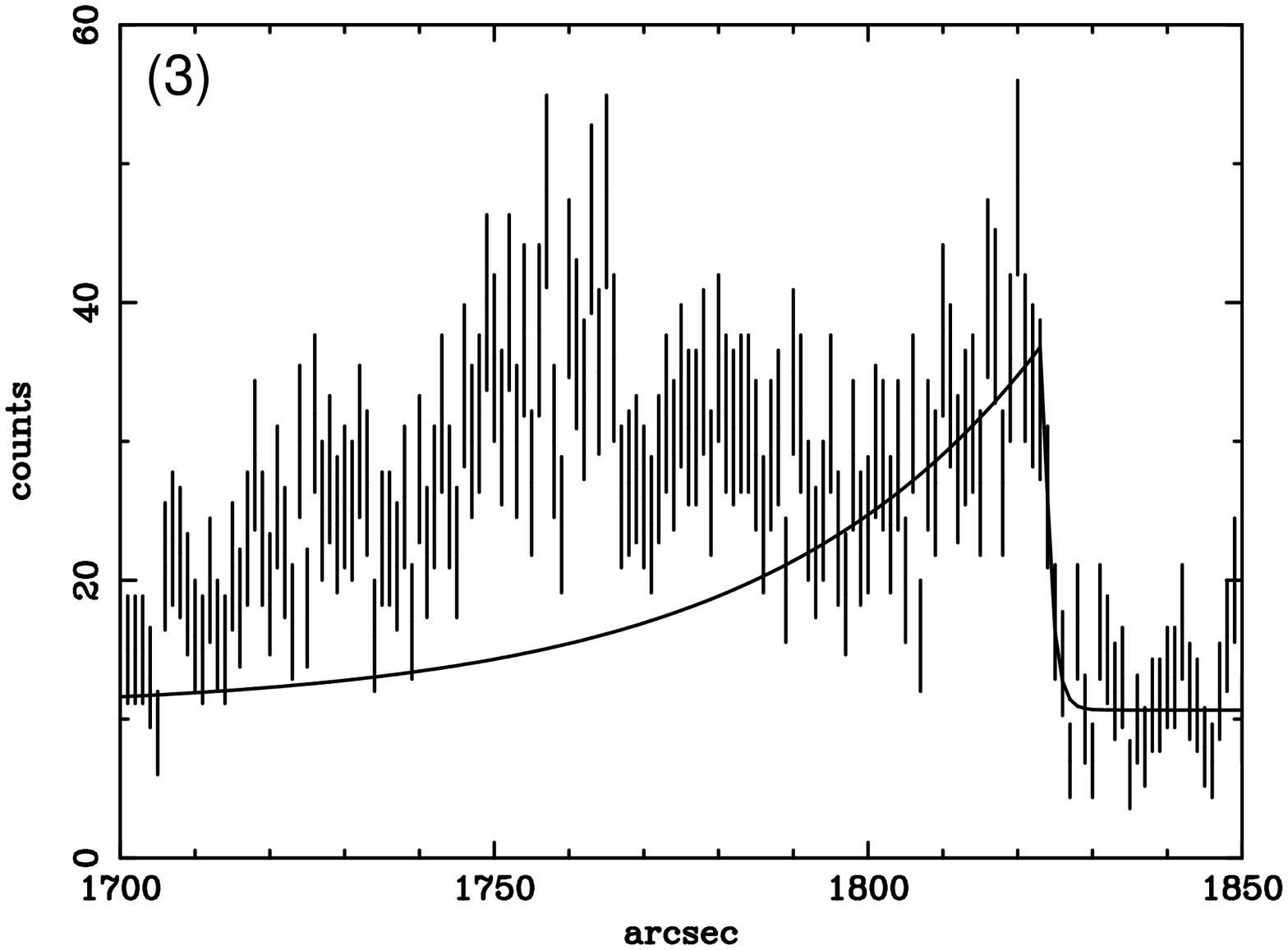}
\caption{Profiles of the filaments in the 2.0--10.0~keV band,
binned with 1~arcsec scale.
The best-fit models are shown with solid lines.
In each panel, the shock runs from left to right.
For filaments~2 and 3,
each fitting was carried out only around the outer peak
in order to avoid the  contamination from the inner peak.
}
\label{fig:profiles}
\end{figure}

The best-fit
values of the photon index and absorption column are
consistent with 
those of the nonthermal component
in previous results \citep{slane2001,iyudin2005,hiraga2005},
indicating that 
almost all photons 
above 2~keV are nonthermal origin.
Thus, we regarded all the photons above 2~keV as nonthermal
in the following.
As shown in Figure~\ref{fig:images}(b), three filaments were selected
(filament 1--3)
in order to study their spatial and spectral characteristics,
which are straight and free from other structures,
in the same way to the analysis of SN~1006 case \citep{bamba2003}.
Since all the filaments are located within 4~arcmin from the aim-point
(see Table~\ref{obs_log}),
the size of 
a point spread function (PSF) 
is about 0.5~arcsec.
Although we can see the filaments more clearly in the soft X-ray band
in Figure~\ref{fig:images},
\citet{bamba2003} showed us that
the contamination of thermal photons makes filaments broader.
Therefore, we conducted the spatial analysis 
only with photons above 2~keV.
Figure~\ref{fig:profiles} shows 
spatial profiles of the filaments
in the 2.0--10.0~keV (nonthermal) band
binned to a resolution of 1~arcsec.
We can see profiles
with clear decays on the downstream sides
and sharp edges on the upstream sides 
in all profiles.
The filament 2 and 3 are double peaked,
as suggested by \citet{iyudin2005}.
To estimate the scale width of these filaments
on upstream ($w_u$) and downstream ($w_d$) sides,
we fitted them with exponential function
as a simple and empirical fitting model
for the apparent profiles,
which is the same
as used in the previous analysis in the SN~1006 \citep{bamba2003}
and other historical SNR \citep{bamba2005} cases.
Background photons were not subtracted
and treated as a 
spatially independent component.
For the filament~3, the $w_u$ could not be determined
due to the lack of statistics,
then it was frozen to be the PSF size (0.5~arcsec).
For each filament, fitting region was limited around the outermost peak
in order to avoid the influence of the inner structure.
The model well reproduced the data.
The best-fit models and parameters are shown in 
Figure~\ref{fig:profiles} and Table~\ref{tab:pro_para}, respectively.

For the next step,
we made spectra of the filaments.
The background regions were selected
from the downstream sides of the filaments,
where there was no other structure.
The spectra were so hard
and fitted with an absorbed power-law model again.
The absorption column was fixed to be $6.1\times 10^{21}~{\rm cm^{-2}}$
in order to make statistics better,
which is the best-fit value for the total region (Table~\ref{total_spec})
and consistent with previous results.
The index of the power-law component is similar to 
those of nonthermal shells in other SNRs
(SN~1006, \citealt{bamba2003};
RX~J1713.7$-$3946, \citealt{koyama1997,slane1999}),
which is considered to be synchrotron emission.
Therefore, we concluded that the nonthermal emission in Vela Jr. shell is
synchrotron,
and applied the {\tt SRCUT} model,
which is one of the model representing the synchrotron emission
from electrons of power-law distribution
with exponential roll-off in a homogeneous magnetic field
\citep{reynolds1998,reynolds1999}.
The spectral index at 1~GHz was fixed to be $0.3$
according to a report by \citet{combi1999}.
Since the value $p=0.3$ is rather smaller than those of ordinary SNRs,
about 0.5,
we also tested the model under the assumption of $p=0.5$.
All models can reproduce the data with 
similar values of reduced $\chi^2$
as shown in Table~\ref{tab:spec_para}.

\section{Comments on $^{44}$Sc fluorescent line}
\label{sec:Sc-line}

\begin{figure}[hbtp]
\epsscale{0.5}
\plotone{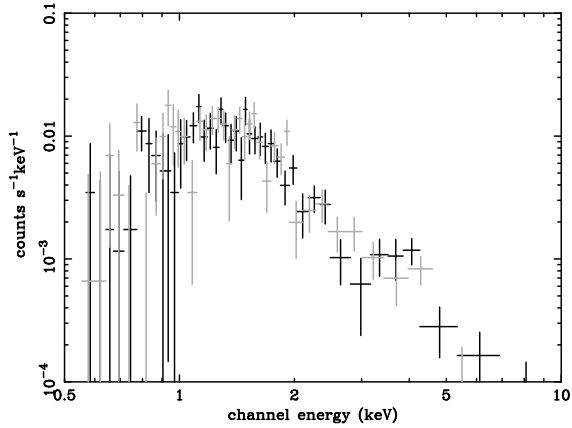}
\caption{
Spectra of the filament 1.
The black and gray data set represent
the data for obsID = 3846 and 4414, respectively.}
\label{fig:4keV-line}
\end{figure}

\citet{iyudin2005} reported a significant emission line
at 4.45$\pm$0.05~keV
from the NW rim of Vela Jr.
with {\it XMM-Newton}.
We checked the spectrum of each filament as well as the averaged one
so as to investigate the presence of some line-like feature.
As a result,
we found that
the spectrum of filament~1 has an excess
around 4~keV as shown in Figure~\ref{fig:4keV-line},
while
the spectrum averaged for whole rim does not show the line feature.
Since one can see the excess in the data of both observations,
it may be a real line.
Then, the power-law plus a narrow line model were applied 
and accepted,
although the reduced $\chi^2$ does not improve significantly
($\chi^2$/d.o.f.=69.5/79; see also Table~\ref{tab:spec_para}).
The center energy is $4.1\pm0.2$~keV,
which is slightly lower than the averaged one by \citet{iyudin2005}
but consistent with the {\it ASCA} result \citep{tsunemi2000}.
The total flux, $7.3_{-4.5}^{+5.1}\times 10^{-7}$~photons~cm$^{-2}$s$^{-1}$,
is about 10\% of that derived  by \citet{iyudin2005}
and \citet{tsunemi2000},
and consistent with the upper limit by \citet{slane2001}.
The difference of the intensity may be because of
our smaller region the spectra were accumulated than
those by other authors.
Moreover, the center energy and significance depend on  the choice of 
the background region and the model of the continuum component.
The lack of statistics 
prevents us from reliable conclusion.
Detailed analysis with excellent statistics and spectral resolution
are encouraged.

\section{Origin of the Filaments}
\label{sec:discuss}

We found that
the scale width of the filaments in Vela Jr. 
is much smaller than its radius.
Filament~3 has a small $w_u$ in the order of {\it Chandra} PSF size
(0.5~arcsec),
indicating that 
the $w_u$ may be similar to or smaller than the PSF size.
Although the other filaments have $w_u$ significantly larger than the PSF size,
it may be caused by the projection effect
or other structures, such as the second peak and/or other filaments.
Therefore,
we concluded that the scale width on the upstream side is
similar to or smaller than the PSF size.
On the other hand,
the value of $w_d$ is much larger than
the PSF size.
These results are similar to other case of nonthermal filaments in young SNRs
\citep{bamba2005}.
The power-law index is similar to the results with previous observations
\citep{slane2001,hiraga2005}.
These thin filaments with hard X-rays remind us
the nonthermal X-rays from filaments in young SNRs.
Wide band spectra of most of these SNRs show that
the nonthermal X-rays are synchrotron radiation
from highly accelerated electrons.
Therefore, we consider that
the filaments emit X-rays 
via synchrotron radiation.
In order to confirm our conclusion,
we need more information in the radio continuum band
to make the wide band spectrum of synchrotron emission.

\section{Interaction with Molecular Clouds?}
\label{sec:MC}

The filaments in Vela Jr. 
are rather straight and their lengths are comparable to
the radius of the SNR, that are similar to
those in SN~1006 \citep{bamba2003} and Tycho \citep{hwang2002},
and are unlike
clumpy filaments in Cas~A \citep{vink2003}, RCW~86 \citep{rho2002},
and RX~J1713.7$-$3946 \citep{uchiyama2003,lazendic2004,cassam-chenai2004}.
The former samples are located in tenuous interstellar space,
whereas for the two SNRs in the latter cases,
there are some reports about 
the interaction between shocks and molecular clouds
(RCW~86: Moriguchi et al., private communication;
RX~J1713.7$-$3946: Fukui et al.\ 2003).
Interactions with molecular clouds may distort
beautiful filaments via turbulence and so on.
There are molecular clouds around Vela Jr. \citep{moriguchi2001}.
However, it is not clear
whether the shocks are interacting with them or not,
because there is no observation with excited molecular cloud lines
(CO(2$\rightarrow$1) and so on).
The straight filaments may indicate that
there is no interaction between the shock and the molecular cloud
in this region.
Further observations with excited molecular lines are needed.

\section{Origin of TeV $\gamma$-rays}
\label{sec:TeV}

Recently, two instruments have independently detected
TeV $\gamma$-rays significantly \citep{katagiri2005,aharonian2005}.
The reported differential flux is consistent with each other,
but the photon index is slightly different.
In this section,
we examine the origin of the TeV $\gamma$-rays,
considering the preferable values of
the maximum energy of electrons ($E_{max}$)
and the downstream magnetic field ($B_d$)
\citep{yamazaki2004}.
\citet{bamba2005} suggested that the width of the filament on the downstream
side ($w_d$) and the roll-off frequency of synchrotron emission
($\nu_{rolloff}$) reflect the value of $E_{max}$ and $B_d$,
then we can use these two observational value for the study on
$E_{max}$ and $B_d$.
Hereinafter, we adopt flux-averaged mean values, such as
\begin{eqnarray}
w_d &=& 49.5~{\rm arcsec}
= 0.24_{-0.07}^{+0.19}\left(\frac{D}{\rm 1~kpc}\right)~{\rm pc}\ \ ,
\label{eq:wd}\\
\nu_{rolloff} &=& 6.6_{-1.6}^{+2.1}\times 10^{16}~{\rm Hz}\ \ ,
\label{eq:nu}
\end{eqnarray}
where $D$ is the distance to Vela Jr.
Considering the projection effect,
$w_d$ can be written as
\begin{eqnarray}
w_d &=& \alpha r^{-1}u_st_{loss}\ \ , \nonumber \\
t_{loss} &=& 1.25\times 10^{3}~{\rm yrs}
\left(\frac{E_{max}}{100~{\rm TeV}}\right)^{-1}
\left(\frac{B_d}{10~{\rm \mu G}}\right)^{-2}\ \ , \nonumber
\end{eqnarray}
where $\alpha$, $r$, $u_s$, and $t_{loss}$ are
a correction factor of the projection effect,
the compression ratio, the shock velocity,
and synchrotron loss time scale
\citep{bamba2005,yamazaki2004}.
Although $\alpha$ is an unknown parameter
depending on the shape of profiles and the curvature radius,
it becomes $\sim$1--10 as far as 
the width of filaments is negligible
to the SNR radii
\citep{berezhko2004}.
Then the above equations lead 
\begin{equation}
\left(\frac{E_{max}}{100~{\rm TeV}}\right)^{-1}
\left(\frac{B_d}{10~{\rm \mu G}}\right)^{-2}
=
6.3_{-1.8}^{+5.0}\times 10^{-2}
\alpha^{-1}r
\left(\frac{u_s}{\rm 3000~km~s^{-1}}\right)^{-1}
\left(\frac{D}{\rm 1~kpc}\right)\ \ .
\label{eq:1}
\end{equation}
On the other hand,
the roll-off frequency $\nu_{rolloff}$ is represented as
\citep{reynolds1999,reynolds1998}\footnote{%
See http://heasac.gsfc.nasa.gov/docs/xanadu/xspec
for the erratum of coefficients.
}
\begin{eqnarray}
\nu_{rolloff} &=& 1.6\times 10^{18}~{\rm Hz}
\left(\frac{E_{max}}{100~{\rm TeV}}\right)^{2}
\left(\frac{B_d}{10~{\rm \mu G}}\right)\ \ .
\label{eq:2}
\end{eqnarray}
In the following, we adopt typical values
$\alpha = 5$ and $r=4$. Then,
using Eqs.~(\ref{eq:1}) and (\ref{eq:2}), 
we derived
\begin{eqnarray}
E_{max} &=& 4.1_{-0.9}^{+0.9}\ {\rm TeV}\
\left(\frac{u_s}{\rm 3000~km~s^{-1}}\right)^{-1/3}
\left(\frac{D}{\rm 1~kpc}\right)^{1/3}
\left(\frac{\alpha}{5}\right)^{-1/3}
\left(\frac{r}{4}\right)^{1/3}
\ \ , 
\label{eq:E}\\
B_d &=& 2.2_{-0.2}^{+0.2}\times10^2\ {\rm \mu G}\
\left(\frac{u_s}{\rm 3000~km~s^{-1}}\right)^{2/3}
\left(\frac{D}{\rm 1~kpc}\right)^{-2/3}
\left(\frac{\alpha}{5}\right)^{2/3}
\left(\frac{r}{4}\right)^{-2/3}
\ \ .
\label{eq:B}
\end{eqnarray}
\citet{katagiri2005} examined whether the TeV $\gamma$-rays arise from the 
inverse Compton emission via accelerated electrons
or $\pi^0$ decay caused by accelerated protons.
The former requires very high size ratio of X-ray and TeV $\gamma$-ray
emission regions ($V_{\rm TeV}/V_{\rm X-ray} \sim 10^5$)
and strong magnetic field of $B_d\sim$1.6~mG.
Then we obtain the shock velocity with eq.(\ref{eq:B}) as 
\begin{equation}
u_s \sim 6\times10^8\ {\rm cm~s^{-1}}\
\left(\frac{D}{\rm 100~pc}\right)
\left(\frac{\alpha}{5}\right)^{-1}
\left(\frac{r}{4}\right) ~~,
\end{equation}
so that the small distance, $D\lesssim100$~pc
(the very small physical radius and the very young age,
in other words)
may be required,
which is somewhat doubtful.
This result is consistent with the discussion in \citet{aharonian2005}.
On the other hand,
$\pi^0$-decay model, which requires the interaction of
the SNR and molecular clouds, can naturally explain
the multi-band spectrum \citep{katagiri2005}. However,
there might be some discrepancy between the observed straight filaments
as discussed in \S~\ref{sec:MC}.

\section{Estimation of the Age and the Distance}
\label{sec:estimation}

\begin{figure}[hbtp]
\epsscale{0.5}
\plotone{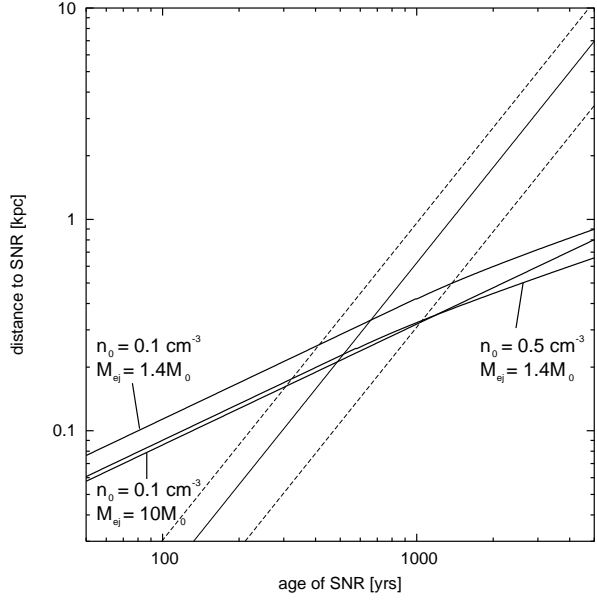}
\caption{The relations between the age and the distance
derived from the evolution of SNRs and from the function ${\cal B}$.
The bold, solid, and dashed lines are
the relations from the SNR model, and best-fit one from the function ${\cal B}$
and their allowed regions, respectively.}
\label{fig:distance}
\end{figure}

More determinative arguments on the distance ($D$) can be possible,
and at this time, the age of the SNR ($t_{age}$) can be also discussed
simultaneously.
As a tool of the estimation,
$\cal B$---age relation \citep{bamba2005} is used,
where 
\begin{eqnarray}
{\cal B} &\equiv& \nu_{rolloff}/w_d{}^2 = C t_{age}{}^{\alpha}\ \ ,\\
C &=& 2.6_{-1.4}^{+1.2}\times 10^{27}\ \ {\rm Hz pc^{-2}}, \\
\alpha &=& -2.96_{-0.06}^{+0.11}\ \ .
\end{eqnarray}
Let $\theta_{R}$ ($\equiv R_s/D$) and $\theta_d$ ($\equiv w_d/D$) be
the angular radius of the SNR and the observed angular scale width
of synchrotron X-ray filaments
on the downstream sides, respectively.
Here $R_s$ is the radius of the SNR in the unit of pc.
From the equation defining $\cal B$, we obtain
\begin{eqnarray}
D &=& 0.65~{\rm kpc}\frac{10~{\rm arcsec}}{\theta_d}
\left(\frac{\nu_{rolloff}}{10^{17}~{\rm Hz}}\right)^{1/2}
\left(\frac{{\cal B}(t_{age})}{\rm 10^{20}~Hz~pc^{-2}}\right)^{-1/2}\ \ .
\label{eq:D-B}
\end{eqnarray}
The thin solid and dashed lines in Figure~\ref{fig:distance}
represent the relation between the age and the distance
using eq.(\ref{eq:D-B}), (\ref{eq:wd}), and (\ref{eq:nu}).

On the other hand, $D=R_s/\theta_R$ is rewritten as
\begin{eqnarray}
D &=& 0.34~{\rm kpc}\frac{10~{\rm arcmin}}{\theta_{R}}
\frac{R_s(t_{age})}{\rm 1~pc}\ \ ,\\
R_s(t_{age}) &=& \left\{
\begin{array}{ll}
1.1~{\rm pc}\left(\frac{E_{51}}{M_{ej}\rho_0}t_{age}{}^4\right)^{1/7} & t<t_{ST} \\
1.2~{\rm pc}\left(\frac{E_{51}}{\rho_0}\right)^{1/5}
\left(t_{age}-0.22E_{51}{}^{-1/2}M_{ej}{}^{5/6}\rho_0{}^{-1/3}
\right)^{2/5}
& t>t_{ST}
\end{array}
\right. \ \ .
\end{eqnarray}
The function $R_s(t_{age})$ is 
given by \citet{truelove1999}
(see eq.(1) and (2), and table~6 and 7),
which studies the shock dynamics of SNRs,
where $E_{51}$, $M_{ej}$, $\rho_0$, and $t_{ST}$ is
the explosion energy in the unit of $10^{51}$~ergs,
the ejecta mass in the unit of $M_\odot$, 
the ambient density in the unit of g~cm$^{-3}$, 
and the time scale that
SNRs enter the Sedov-Taylor phase
(in which shock begins to deaccelerate due to the interstellar medium),
respectively \citep{truelove1999}.
For simplicity,
we consider the case of the constant interstellar medium, $n_0$,
\begin{equation}
n_0 \equiv \frac{\rho_0}{\mu_H}\ \, \nonumber
\end{equation}
where $\mu_H$ is the mean mass per hydrogen nucleus,
$1.4\times 1.67\times 10^{-24}$~g.
In the following,
we choose the kinetic energy of the ejecta, the ejecta mass,
and $n_0$ 
to be $10^{51}$~ergs, 1.4$M_\odot$, and 0.1~cm$^{-3}$,
respectively.
Solving these equations, one can estimate both $D$ and $t_{age}$.

Figure~\ref{fig:distance} represents 
the relation between the age and the distance.
We derived the allowed range of the parameters to be
\begin{eqnarray}
t_{age} &=& 660\ (420 - 1400)\ {\rm yrs}\ \ ,\\
D &=& 0.33\ (0.26-0.50)\ {\rm kpc}\ \ .
\end{eqnarray}
When we vary the ambient number density into $n_0 = 0.5~{\rm cm^{-3}}$
or the ejecta mass into $M_{ej} = 10~M_\odot$,
the result remains basically unchanged as can be seen 
in Figure~\ref{fig:distance}.

Although the allowed regions are too wide
to derive some conclusion,
we may be able to say that Vela Jr. is a nearby and relatively young SNR.
The results are consistent with previous reports,
which may imply
that this is an indirect confirmation of $\cal B$---age relation 
\citep{bamba2005}
and our distance/age indicator may become a useful tool in the future.
Furthermore, derived distance is significantly larger than that 
of Vela SNR \citep[250~pc;][]{cha1999},
then there may be no interaction between these two SNRs.

The most common distance indicator of SNR is
an application of
the relation between the surface brightnesses at 1~GHz and the diameters
of SNRs \citep[$\Sigma$---$D$ relation;][]{case1998},
however, the relation has still large uncertainties.
Recent X-ray surveys recognize
a new type of SNRs,
which are dim in radio band and bright in hard X-ray band,
such as RX~J1713.7$-$3946 \citep{koyama1997},
G28.6$-$0.1 \citep{bamba2001,ueno2003},
and so on \citep[e.g.][]{combi2005,yamaguchi2004}.
The number of such SNRs may be more than $\sim$20
\citep{bamba2003a}.
They must be significant accelerators of cosmic rays
because most of them emit synchrotron X-rays,
Unfortunately, they are dim in radio bands, so that
the $\Sigma$---$D$ relation cannot directly applied.
Our method may also be an useful tool to estimate both $D$ and $t_{age}$
of such synchrotron X-ray emitting SNRs.

Again let us consider $E_{max}$ and $B_d$
using Eqs.~(\ref{eq:E}) and (\ref{eq:B}).
Assuming that $D\sim300~\rm pc$, that is the most preferable value
of our distance indicator,
we obtain
$E_{max}\sim 3$~TeV and $B_d \sim 5\times 10^2~\mu$G.
Although the uncertainty is very large,
our result may indicate that
the magnetic field is highly amplified.

These results does not include the uncertainty of 
the model of SNR dynamics and the function ${\cal B}$.
Considered these uncertainty,
the allowed regions for $t_{age}$ and $D$ become more larger.
It is the uncertainty of $\cal B$'s normalization
which makes error regions the most widest,
about 50\%.
In order to improve this method for $t_{age}$ and $D$,
more samples are needed
to make the uncertainty of $\cal B$ smaller.

\section{Summary}
\label{sec:summary}

We have conducted systematic spectral and spatial analysis of
filamentary structures
in Vela Jr.\ NW rim for the first time.
A summary of our results is as follows:

\begin{enumerate}
\item
We found that
nonthermal X-rays from Vela Jr. NW rim are concentrated on
very thin filamentary structures.
The average scale width on the upstream side is
similar to or smaller than
the PSF size of {\it Chandra},
whereas that on the downstream side is 
49.5 (36.0--88.8) arcsec.
\item
The spectra of filaments are hard and have no line-like structure,
which is well reproduced with an absorbed power-law model of
$\Gamma$ = 2.67 (2.55--2.77),
or {\tt SRCUT} model with $\nu_{rolloff}$ = 4.3 (3.4--5.3)$\times 10^{16}$~Hz
under the assumption of $p=0.3$.
\item
We tried to estimate the distance $(D)$ and the age $(t_{age})$ of Vela Jr.
using the function $\cal B$
and estimated that
$t_{age}$ = 660\ (420--1400)~yrs and $D$ = 0.33 (0.26--0.50)~kpc,
which is consistent with previous reports.
These results may suggest that
there is no interaction between Vela SNR and Vela Jr.
\item
Using the estimated $D$,
we derived 
the most preferable values
$E_{max}\sim 3~{\rm TeV}$ and
$B_d\sim 500~\mu$G.
Our result may imply that
the magnetic field on the filament is highly amplified.
\end{enumerate}

\acknowledgements

Our particular thanks are due to
the anonymous referee,
K. Makishima, F. Takahara, Y. Mochizuki, Y. Moriguchi, 
Y. Uchiyama, M. Tsujimoto, J. Vink, and K. Ebisawa,
for their fruitful discussions and comments.
R.Y. and J.S.H. are supported by JSPS Research Fellowship for Young Scientists.

\onecolumn

\begin{deluxetable}{p{7pc}cccc}
\tabletypesize{\scriptsize}
\tablecaption{Observation Log
\label{obs_log}}
\tablewidth{0pt}
\tablecolumns{2}
\tablehead{
 \colhead{ObsID} & \colhead{R.A.} &
\colhead{Dec.} & \colhead{Date} & \colhead{Exposure (ks)} 
}
\startdata
3846\dotfill & 08\fh49\fm09\fs3 & -45\fd37\fm42\fs4 & 2003 Jan. 5 & 39 \\
4414\dotfill & 08\fd49\fm09\fs3 & -45\fd37\fm42\fs3 & 2003 Jan. 6 & 35 
\enddata
\end{deluxetable}

\begin{deluxetable}{p{9pc}c}
\tabletypesize{\scriptsize}
\tablecaption{%
Best-fit parameters of the Spectrum of the NW rim%
\tablenotemark{a}
\label{total_spec}}
\tablewidth{0pt}
\tablecolumns{2}
\tablehead{
\colhead{Parameters} & \colhead{Best-Fit Value} 
}
\startdata
$\Gamma$\dotfill & 2.67 (2.52--2.84) \\
$N_{\rm H}$ [$10^{21}$~H~cm$^{-2}$]\tablenotemark{b}\dotfill & 6.1 (2.4--9.9) \\
Flux\tablenotemark{c} [$10^{-12}$ergs~cm$^{-2}$s$^{-1}$]\dotfill & 6.2 (5.9--6.5) \\
$\chi^2$/d.o.f.\dotfill & 114.4/135 \\
\enddata
\tablenotetext{a}
{Parentheses indicate single parameter 90\% confidence regions.}
\tablenotetext{b}
{Calculated using the cross sections by \citet{morrison1983}
with the solar abundances \citep{anders1989}.}
\tablenotetext{c}
{In the 2.0--10.0~keV band.}
\end{deluxetable}

\begin{deluxetable}{p{3pc}cccc}
\tabletypesize{\scriptsize}
\tablecaption{Best-fit parameters of the profiles of the filaments.%
\tablenotemark{a}
\label{tab:pro_para}}
\tablewidth{0pt}
\tablecolumns{2}
\tablehead{
\colhead{No.} & \colhead{$A$} & \colhead{$w_u$}
 & \colhead{$w_d$} & \colhead{Reduced $\chi^2$} \\
 & [cnts arcsec$^{-1}$] & [arcsec] &  [arcsec] & [$\chi^2$/d.o.f.]
}
\startdata
1\dotfill & 17.8 (15.8--19.9) & 19.0 (12.1--31.4) & 31.8 (23.7--47.1) & 143.4/121 \\
2\dotfill & 41.4 (36.5--47.4) & 3.68 (2.69--5.75) & 65.0 (38.2--144.6) & 56/51 \\
3\dotfill & 26.4 (22.4--30.7) & \nodata\tablenotemark{b} & 37.1 (26.5--60.2) & 58/54 \\
mean\tablenotemark{c}\dotfill & \nodata & \nodata & 49.5 (34.8--98.3) & \nodata 
\enddata
\tablenotetext{a}
{Parentheses indicate single parameter 90\% confidence regions.}
\tablenotetext{b}
{Fixed to be 1~arcsec.}
\tablenotetext{c}
{Flux-weighted mean value.}
\end{deluxetable}

\begin{deluxetable}{p{2pc}ccccccccc}
\tabletypesize{\scriptsize}
\tablecaption{Best-fit parameters of the spectra of the filaments.%
\tablenotemark{a}
\label{tab:spec_para}}
\tablewidth{0pt}
\tablecolumns{2}
\tablehead{
 & \multicolumn{3}{c}{\tt power-law} & \multicolumn{3}{c}{{\tt srcut} ($p=0.3$)} & \multicolumn{3}{c}{{\tt srcut} ($p=0.5$)} \\
\cline{2-4} \cline{5-7} \cline{8-10}
\colhead{No.} & \colhead{$\Gamma$} & \colhead{Flux\tablenotemark{b}} & \colhead{$\chi^2$/d.o.f.} & \colhead{$\nu_{rolloff}$\tablenotemark{c}} & \colhead{$\Sigma_{\rm 1GHz}$\tablenotemark{d}} & \colhead{$\chi^2$/d.o.f.} & \colhead{$\nu_{rolloff}$\tablenotemark{c}} & \colhead{$\Sigma_{\rm 1GHz}$\tablenotemark{d}} & \colhead{$\chi^2$/d.o.f.}
}
\startdata
1\dotfill & 2.87 & 0.74 & 76.2/81 & 2.7 & 4.4 & 79.2/81 & 3.7 & 148 & 76.1/81 \\
 & (2.71--3.06) & (0.63--0.85) & & (1.8--3.8) & (2.8--3.8) & & (2.6--5.8) & (95--230) & \\
2\dotfill & 2.59 & 1.8 & 137.3/118 & 5.3 & 1.2 & 141.3/118 & 8.2 & 120 & 141.3/118 \\
 & (2.44--2.74) & (1.6--2.1) & & (3.6--7.3) & (1.1--1.3) & & (5.5--12.8) & (113--127) & \\
3\dotfill & 2.45 & 0.76 & 56.2/44 & 7.4 & 0.93 & 57.7/44 & 11.6 & 32 & 57.4/44 \\
 & (2.23--2.68) & (0.61--0.92) & & (4.1--13.2) & (0.84--1.01) & & (6.4--25.7) & (28--36) & \\
Total\dotfill & 2.67 & 3.2 & 276.8/245 & 4.3 & 8.8 & 286.6/245 & 6.6 & 282 & 284.9/245 \\
 & (2.57--2.77) & (2.9--3.6) & & (3.4--5.3) & (8.1--9.5) & & (5.0--8.7) & (262--301) & 
\enddata
\tablecomments{
The absorption column was fixed to be $6.1\times 10^{21}$~H~cm$^{-2}$
according to Table~\ref{total_spec}.
}
\tablenotetext{a}
{Parentheses indicate single parameter 90\% confidence regions.}
\tablenotetext{b}
{Absorption-corrected flux in the 2.0--10.0~keV band in the unit of $10^{-13}$~ergs~cm$^{-2}$s$^{-1}$.}
\tablenotetext{c}
{Roll-off frequency in the unit of $10^{16}$~Hz.}
\tablenotetext{d}
{Flux density at 1~GHz in the unit of $10^{-4}$~Jy.}
\end{deluxetable}


\begin{thebibliography}{}
\bibitem[Aharonian et al.(2005)]{aharonian2005}
Aharonian et al.\ 2005, accepted by \aap letters
(astroph/0505380)
\bibitem[Anders \& Grevesse(1989)]{anders1989}
Anders, E., \& Grevesse, N. 1989, \gca, 53, 197
\bibitem[Aschenbach(1998)]{achenbach1998}
Aschenbach, B.\ 1998, \nat, 396, 141
\bibitem[Bamba et al.(2000)Bamba, Tomida, \& Koyama]{bamba2000}
Bamba, A., Tomida, H., \& Koyama, K. 2000, \pasj, 52, 1157
\bibitem[Bamba et al.(2001)]{bamba2001}
Bamba, A., Ueno, M., 
Koyama, K., \& Yamauchi, S.\ 2001, \pasj, 53, L21
\bibitem[Bamba et al.(2003a)]{bamba2003a}
Bamba, A., Ueno, M., 
Koyama, K., \& Yamauchi, S.\ 2003a, \apj, 589, 253
\bibitem[Bamba et al.(2003b)]{bamba2003} 
Bamba, A., Yamazaki, R., Ueno, M., \& Koyama, K.\ 2003b, \apj, 589, 827 
\bibitem[Bamba(2004)]{bambaphd}
Bamba, A.\ 2004, Ph.D.~thesis (Kyoto University)
\bibitem[Bamba et al.(2005)]{bamba2005}
Bamba, A., Yamazaki, R., Yoshida, T., Terasawa, T., \& Koyama, K.\ 2005, 
\apj, 621, 793
\bibitem[Bell(1978)]{bell1978}
Bell, A.~R.\ 1978, \mnras, 182, 443
\bibitem[Bell \& Lucek(2001)]{bell2001}
Bell, A.~R.~\& Lucek, S.~G.\ 2001, \mnras, 321, 433
\bibitem[Berezhko \& V{\"o}lk(2004)]{berezhko2004}
Berezhko, E.G., \& V{\"o}lk, H.J,\ 2004, \aap, 419, L27 
\bibitem[Blandford \& Eichler(1987)]{blandford1987}
Blandford, R.~D., \& Eichler, D.\ 1987, \physrep, 154,1
\bibitem[Borkowski et al.(2001)]{borkowski2001}
Borkowski, K.~J., Rho, J., Reynolds, S.~P. \& Dyer, K.~K. 2001, \apj, 550, 334
\bibitem[Blandford \& Ostriker(1978)]{blandford1978}
Blandford, R.~D., \& Ostriker, J.~P.\ 1978, \apj, 221, L29
\bibitem[Burgess \& Zuber(2000)]{burgess2000}
Burgess, C.~P.~\& Zuber, K.\ 2000, Astroparticle Physics, 14, 1
\bibitem[Case \& Bhattacharya(1998)]{case1998}
Case, G.~L., \& Bhattacharya, D.\ 1998, \apj, 504, 761 
\bibitem[Cassam-Chena{\" i} et al.(2004)]{cassam-chenai2004} 
Cassam-Chena{\" i}, G., Decourchelle, A., Ballet, J., Sauvageot, J.-L., 
Dubner, G., \& Giacani, E.\ 2004, \aap, 427, 199
\bibitem[Cha et al.(1999)]{cha1999}
Cha, A.~N., Sembach, K.~R., \& Danks, A.~C.\ 1999, \apjl, 515, L25
\bibitem[Combi, Romero, \& Benaglia(1999)]{combi1999}
Combi, J.~A., Romero, G.~E., \& Benaglia, P.\ 1999, \apjl, 519, L177
\bibitem[Combi et al.(2005)]{combi2005}
Combi, J.A., Benaglia, P., Romero, G.E., \& Sugizaki, M.\ 2005, 
\aap, in press (astroph/0501051)
\bibitem[Drury(1983)]{drury1983}
Drury, L.O'C.\ 1983, Rep. Prog. Phys., 46, 973
\bibitem[Fukui et al.(2003)]{fukui2003}
Fukui, Y., et al.\ 2003, \pasj, 55, L61
\bibitem[Garmire et al.(2000)]{garmire2000}
Garmire, G., Feigelson, E.~D., Broos, P., Hillenbrand, L.~A.,
Pravdo, S.~H., Townsley, L.,\&  Tsuboi, Y.\ 2000, \aj, 120, 1426
\bibitem[Green(2004)]{green2004}
Green, D.~A.\ 2004,
A Catalogue of Galactic Supernova Remnants (2004 January version), 
(Cambridge, UK, Mullard Radio Astronomy Observatory)
available on the WWW at http://www.mrao.cam.ac.uk/surveys/snrs/
\bibitem[Hiraga et al.(2005)]{hiraga2005}
Hiraga, J.~S., et al.\ 2005, \apj, submitted
\bibitem[Hwang et al.(2002)]{hwang2002} 
Hwang, U., Decourchelle, A., Holt, S.~S., \& Petre, R.\ 2002, \apj, 581, 1101 
\bibitem[Iyudin et al.(1998)]{iyudin1998}
Iyudin, A.~F., et al.\ 1998, \nat, 396, 142 
\bibitem[Iyudin et al.(2005)]{iyudin2005}
Iyudin, A.~F., Aschenbachm B., Becker, W., Dennerl, K., \& Haberl, F.\ 2005,
\aap, 429, 225
\bibitem[Jones \& Ellison(1991)]{jones1991}
Jones, F.C., \& Ellison, D.C.\ 1991, Space Science Rev., 58, 259
\apj, 580, 1060
\bibitem[Katagiri et al.(2005)]{katagiri2005}
Katagiri, H. et al.\ 2005, \apjl, in press (astro-ph/0412623)
\bibitem[Koyama et al.(1997)]{koyama1997}
Koyama, K., Kinugasa, 
K., Matsuzaki, K., Nishiuchi, M., Sugizaki, M., Torii, K., Yamauchi, S., \& 
Aschenbach, B.\ 1997, \pasj, 49, L7
\bibitem[Koyama et al.(1995)]{koyama1995}
Koyama, K., Petre, R., Gotthelf, E.V., Hwang, U., Matsura, M., 
Ozaki, M., \& Holt S.~S.\ 1995, \nat, 378, 255
\bibitem[Lazendic et al.(2004)]{lazendic2004}
Lazendic, J.~S., Slane, P.~O., Gaensler, B.~M., Reynolds, S.~P.,
Plucinsky, P.~P., \& Hughes, J.~P.\ 2004, \apj, 602, 271
\bibitem[Long et al.(2003)]{long2003}
Long, K.~S., Reynolds, S.~P., Raymond, J.~C., Winkler, P.~F., Dyer, K.~K.,
\& Petre, R.\ 2003, \apj, 586, 1162
\bibitem[Lucek \& Bell(2000)]{lucek2000}
Lucek, S.~G.~\& Bell, A.~R.\ 2000, \mnras, 314, 65 
\bibitem[Malkov \& Drury(2001)]{malkov2001}
Malkov, E., \& Drury, L.O'C.\ 2001, Rep.\ Prog.\ Phys., 64, 429
\bibitem[Mereghetti, Pellizzoni, \& de Luca(2002)]{mereghetti2002} 
Mereghetti, S., Pellizzoni, A., \& de Luca, A.\ 2002, ASP Conf.~Ser.~271: 
Neutron Stars in Supernova Remnants, 289
\bibitem[Moriguchi et al.(2001)]{moriguchi2001}
Moriguchi, Y., Yamaguchi, N., Onishi, T., Mizuno, A., \& Fukui, Y.\ 
2001, \pasj, 53, 1025
\bibitem[Morrison \& McCammon(1983)]{morrison1983}
Morrison, R., \& McCammon, D.\ 1983, \apj, 270, 119
\bibitem[Pavlov et al.(2001)]{pavlov2001}
Pavlov, G.~G., Sanwal, D., 
K{\i}z{\i}ltan, B., \& Garmire, G.~P.\ 2001, \apjl, 559, L131 
\bibitem[Reynolds(1998)]{reynolds1998}
Reynolds, S.~P.\ 1998, \apj, 493, 375
\bibitem[Reynolds \& Keohane(1999)]{reynolds1999}
Reynolds, S.P., \& Keohane, J.W.\ 1999, \apj, 525, 368
\bibitem[Rho et al.(2002)]{rho2002}
Rho, J., Dyer, K.~K., Borkowski, K.~J., \& Reynolds, S.~P.\ 2002, \apj, 581, 
1116 
\bibitem[Shonfelder et al.(2000)]{shonfelder2000}
Shonfelder, V., et al. 2000, in AIP Conf. Proc. 510, Fifth Compton
Symp., ed. M. L. McConnell \& J. M. Ryan (New York: AIP), 54
\bibitem[Slane et al.(1999)]{slane1999}
Slane, P., Gaensler, 
B.~M., Dame, T.~M., Hughes, J.~P., Plucinsky, P.~P., \& Green, A.\ 1999, 
\apj, 525, 357
\bibitem[Slane et al.(2001)]{slane2001}
Slane, P., Hughes, J.~P., 
Edgar, R.~J., Plucinsky, P.~P., Miyata, E., Tsunemi, H., \& Aschenbach, B.\ 
2001, \apj, 548, 814
\bibitem[Truelove \& McKee(1999)]{truelove1999}
Truelove, J.~K., \& McKee, C.~F.\ 1999, \apjs, 120, 299
\bibitem[Tsunemi et al.(2000)]{tsunemi2000}
Tsunemi, H., Miyata, E., Aschenbach, B., Hiraga, J., \& Akutsu, D.\ 
2000, \pasj, 52, 887
\bibitem[Uchiyama et al.(2003)]{uchiyama2003}
Uchiyama, Y., Aharonian, F.~A., \& Takahashi, T.\ 2003, \aap, 400, 567
\bibitem[Ueno et al.(2003)]{ueno2003}
Ueno, M., Bamba, A., 
Koyama, K., \& Ebisawa, K.\ 2003, \apj, 588, 338
\bibitem[Vink \& Laming(2003)]{vink2003}
Vink, J.~\& Laming, J.~M.\ 2003, \apj, 584, 758 
\bibitem[von Keinlin et al.(2004)]{vonkeinlin2004}
von Keinlin, A., Att{\' e}, D., Schanne, S., Cordier, B., 
Diehl, R., Iyudin, A.~F., Lichti, G.~G., Roques, J.~-P., 
Sch{\" o}nfelder, V., \& Strong, A.\ 2004, 
Proc.~ of the 5th {\it INTEGRAL} Science Workshop,
ESA SP-552 (astrpph/0407129)
\bibitem[Weisskopf et al.(2002)]{weisskopf2002}
Weisskopf, M.~C.,
Brinkman, B., Canizares, C., Garmire, G., Murray, S., \& Van Speybroeck,
L.~P.\ 2002, \pasp, 114, 1
\bibitem[Yamaguchi et al.(2004)]{yamaguchi2004}
Yamaguchi, H., Ueno, 
M., Koyama, K., Bamba, A., \& Yamauchi, S.\ 2004, \pasj, 56, 1059
\bibitem[Yamazaki et al.(2004)]{yamazaki2004}
Yamazaki, R., Yoshida, 
T., Terasawa, T., Bamba, A., \& Koyama, K.\ 2004, \aap, 416, 595
\end{thebibliography}
\end{document}